\documentclass[%
 aip,
 jmp,%
 amsmath,amssymb,
reprint,%
]{revtex4-1}

\usepackage{amssymb}
\usepackage{amsmath}
\usepackage{subfigure}
\usepackage{multirow}
\usepackage{graphicx}
\usepackage{dcolumn}
\usepackage{bm}
\usepackage{color}
\usepackage[table]{xcolor}

\begin{document}


\title[Automatic Generation of Adaptive Network Models based on Similarity to the Desired Complex Network]{Automatic Generation of Adaptive Network Models based on Similarity to the Desired Complex Network}

\author{Niousha Attar}
\email{n\_attar@sbu.ac.ir}
\affiliation{Faculty of Computer Science and Engineering, Shahid Beheshti University G.C, Tehran, Iran}

\author{Sadegh Aliakbary}	
\email{s\_aliakbary@sbu.ac.ir}
\affiliation{Faculty of Computer Science and Engineering, Shahid Beheshti University G.C, Tehran, Iran}

\date{\today}

\begin{abstract}
Complex networks have become powerful mechanisms for studying a variety of real-world systems. Consequently, many human-designed network models are proposed that reproduce nontrivial properties of complex networks, such as long-tail degree distribution or high clustering coefficient. Therefore, we may utilize network models in order to generate graphs similar to desired networks. However, a desired network structure may deviate from emerging structure of any generative model, because no selected single model may support all the needed properties of the target graph and instead, each network model reflects a subset of the required features. In contrast to the classical approach of network modeling, an appropriate modern network model should adapt the desired features of the target network. In this paper, we propose an automatic approach for constructing network models that are adapted to the desired network features. We employ Genetic Algorithms in order to evolve network models based on the characteristics of the target networks. The experimental evaluations show that our proposed framework, called NetMix, results network models that outperform baseline models according to the compliance with the desired features of the target networks.
\end{abstract}

\pacs{}
\keywords{Complex Network, Network Model, Automatic Model Construction, Model Composition, Genetic Algorithm, Social Networks}
\maketitle



\section{Introduction}
\label{sec_intro}
Real-world networks exhibit surprising features that are missing in random or regular graphs. Consequently, various network models are proposed in the literature in order to generate artificial graphs which are topologically similar to real networks. In recent years, network models have found many applications in hypothesis testing, network simulations, and what-if scenarios \cite{arora2017action,bailey2014genetic,modelfit2,you2018graphrnn}. In the classical approach of network modeling, various human-designed models are proposed, each of which follows a specific network formation process and therefore, provides a specific set of network features. Despite the advances in network modeling, a classical network model inherently supports some features and neglects others. For example, Barab\'{a}si-Albert model\cite{BAModel} generates scale-free networks with long-tail degree distribution, but it does not support high clustering. On the other hand, Watts-Strogatz model\cite{WattsStrogatz} generates networks with high clustering but it does not generate scale-free networks. Even the recent human-developed network models fail to support some of the needed network features. Consequently, a network model that is appropriate for a particular target network (e.g., a social network) may fail to synthesize networks similar to another network (e.g., a biological graph). Actually, a human-designed network model controls network formation in a specific manner and therefore, such models are all inherently limited by their specific manners. Consequently, new generators must be developed manually in order to support new demands for network features \cite{arora2017action}. These limitations of the existing network generators reveals the need to a dynamic and adaptive network model which is not restricted to a specific type of network topology or formation process.

When a target network or the set of its desired features is specified, network modeling becomes a complicated task because the model should adapt the target graph, and generate networks compatible with the desired characteristics. Therefore, a research problem has emerged in recent years, which investigates developing network models that are adaptive to the characteristics of the desired network. Adaptive network generators may play role in extrapolation (synthesize larger networks to predict future topology of the network), sampling (synthesize smaller but similar networks), anonymization (synthesize similar networks to a private network), and many other applications \cite{arora2017action}.

In this paper, we develop a method for automatic model construction based on the probabilistic combination of several network generation methods. Actually, we automatically create a network model per target graph based on the topological characteristics of this graph. For any target network, our proposed framework combines existing network generation processes and automatically builds a new mixture model adapted to the target graph. While each network model supports specific set of features, an intelligent combination of the models may lead to adaptive network structures which are more similar to the target graphs. In our proposed framework, all the candidate network generation processes may contribute in network generation, each of which with an assigned probability that is adapted based on the characteristics of the target network. However, finding the best adapted configuration for the combination of the network processes is not a trivial task. We employ Genetic Algorithm in order to find the best probability values assigned to each network process along with their corresponding configuration parameters. Our experiments show that the proposed method outperforms the baseline methods according to the similarity of the synthesized networks to the target graphs. 
 
The rest of this paper is organized as follows: Section \ref{sec:relatedworks} reviews the state of the art network generation methods. In section \ref{sec:problemStatement} the problem statement is presented. Section \ref{sec:proposedmethod} illustrates our proposed method. Section \ref{sec:evaluations} shows the experimental evaluations, including case studies for real-world networks. Finally, Section \ref{sec:conclusion} concludes the paper and explains the future works.

\section{Literature Review} \label{sec:relatedworks}
Many efforts exist in the literature for generating artificial complex networks. Particularly, many algorithms, called network models, are proposed in order to generate graphs with nontrivial topological features of real networks, such as long-tail degree distribution and small-worldness. The classical approach of network modeling is based on reality-inspired but human-designed methods, each of which follows a certain formation mechanism, and supports a specific and preset set of network properties. This approach includes methods such as Erd\H{o}s-R\'{e}nyi model of random graphs \cite{CentralLimit}, Bar\'{a}basi-Albert \cite{BAModel} model of scale-free networks, and Watts-Strogatz (WS) model of Small-World networks \cite{WattsStrogatz}, along with other models such as Kronecker graphs \cite{kronecker}, Random power-law \cite{RandomPowerLaw}, Forest Fire \cite{graphsOverTime}, and several other network growth methods, many of which result in scale-free networks \cite{WebAsGraph, bell2017network, golosovsky2018mechanisms, topirceanu2018weighted}. 

In many applications, a target network or the set of its topological properties is given, and the problem is to generate artificial graphs with features similar to those of the target network. Although network models seem to solve this problem, but each generative model supports a set of network features and ignores other features and consequently, the model is appropriate for a set of target networks and inappropriate for the others \cite{modelfit2}. However, research on developing human-designed models and network formation mechanisms is still ongoing because of its valid applications. For instance, a mechanism of rewiring for tuning two-node degree-correlation and clustering coefficient \cite{kashyap2017mechanisms}, a framework which synthesizes dense and scale-free networks \cite{courtney2018dense}, and a method for generating disassortative graphs with a given degree distribution \cite{van2018generating} are proposed in recent years. 

Some researchers have proposed to select an appropriate model based on the properties of each target network. Particularly, various automatic ``model selection'' methods are proposed in the literature \cite{Drosophila,ModelSelection,GDD,RGF,GraphCrunch2,gmscn,modelfit}, many of which employ machine learning algorithms in order to automatically classify the target network into one of the candidate network models. In this context, decision tree learning \cite{Drosophila,ModelSelection,gmscn} and distance-based classification \cite{GDD,RGF,netdistance,modelfit2,alignmentfreenetcompare,revealHidLang} are frequently utilized. 

Despite the advances in network modeling, the target network is usually a real-world graph with an structure which may deviate from emerging structure of any generative model \cite{medland2016automatic, modelfit2}. Therefore, even with the aid of intelligent model selection methods, no selected single network model may support various properties of the target real-world graph and instead, each network model may reflect a subset of the required features. Although network models are tunable via their configuration parameters, the overall network formation in a model conforms a constant and static process. Consequently, researches have been started to combine different network methods in order to create network models that adaptively imitate the features of a desired network \cite{arora2017action, harrison2015meta, kashirin2016evolutionary, menezes2014symbolic, bailey2014genetic, medland2016automatic, pope2016evolving, verstraaten2016synthetic, bach2012interactive}. Existing works utilize different approaches such as ``Genetic Programming'' (GP) \cite{bailey2014genetic, medland2016automatic, harrison2015meta, verstraaten2016synthetic}, ``Simulated Annealing'' \cite{kashirin2016evolutionary}, cellular automata \cite{lukeman2010inferring}, symbolic regression \cite{menezes2014symbolic}, and dk-graphs \cite{orsini2015quantifying}. For example, Bailey et al. \cite{bailey2014genetic} employ genetic programming (GP) to evolve an algorithmic description of the formation of a target network, but their method is computationally expensive and scales poorly with network size. Recently, an action-based network generator (called ABNG) \cite{arora2017action} is also proposed which orchestrates simple actions of network formation and chooses action probabilities based on simulated annealing to simulate the local interactions of the networks. This method is shown to be effective in simulating different artificial and real networks, and it is regarded as one of the main baselines in evaluating our proposed method. It is also worth noting that in some recent researches, the graph structure is directly utilized (without a feature extraction phase) in order to automatically develop an adaptive network generator \cite{kipf2016variational, Grover2017Graphite, Bojchevski2018NetGan, you2018graphrnn}.

In this work, we propose an extensible framework of network process composition, in which candidate network processes are combined in a probabilistic framework. In comparison to the existing evolutionary methods, our proposed method benefits from the novel approach of network process composition, simple architecture, reproducible implementation, and the ability of extension with new candidate processes.

\section{Problem Statement}
\label{sec:problemStatement}
Synthesized networks represent complex systems in the form of nodes and edges. In a real network, the nodes are connected according to a meaningful and specific pattern \cite{NewmanComplexNets} and therefore, the corresponding synthesized network should conform to a similar formation process. The more a synthesized network is similar to the target network, the more accurate is the result of different experiments on the synthesized network, such as simulation and hypothesis testing. 

Suppose that we can quantify the similarity of two graphs $G_{t}$ and $G_{r}$ via different criteria. If $G_{t}$ is the target network, then the problem is to find a synthesized network $G_{r}$ with maximum similarity to $G_{t}$. Therefore, finding the best $G_{r}$ is a search problem in the extra-large search space of possible graph instances. Moreover, based on the application of network generation, the desired size of the synthesized network may be different from the size of the target network. For example, in an extrapolation application, the synthesized graph is larger than the original target graph. On the other hand, a network sampling application seeks for artificial graphs smaller than the target graph. Consequently, the employed network similarity function that quantifies the similarity of the target network to the artificial graphs, should be a size-independent measure, capable of comparing the topology of two networks with different sizes (e.g., a large and a small graph). 

The ultimate goal in network generation problem is to synthesize graphs matching the desired topological properties of the target network. Consequently, we first need to specify a set of required network properties to be resembled in the synthesized graphs. Then, we should define a network similarity function, as a quality of fit measure which checks the set of required network properties, and quantifies the overall similarity of an artificial network to the target graph. Finally, we should optimize the generator parameters so that the generated graphs become more similar to the target network according the defined similarity measure.

We assume that the considered networks are simple (undirected and unweighted) graphs so that to keep the problem simple and the solution comparable to the main baselines. In summary, a potential solution to the defined problem takes a network (a simple graph) as the input, and generates another simple graph as the output with similar topological characteristics to the inputted complex network, but perhaps with a different (arbitrary) size. 

\section{Proposed Method} \label{sec:proposedmethod}
\subsection{Solution Encoding} \label{sub:solutionencoding}
Suppose that we have a target network $G_{t}$ in hand, and we want to generate an artificial graph $G_{r}$ which is similar to $G_{t}$ regarding its topological features. The amount of dissimilarity of $G_{t}$ and $G_{r}$, called $error(G_{t} , G_{r})$, is measurable via different network properties such as average clustering coefficient, degree distribution, etc. Additionally, suppose that there are several candidate network formation processes C (such as preferential attachment) which are applicable in network generation. In each step of network generation, we select and employ one of the candidate processes $c_i$ with probability $p_i$. For example, we may generate a graph in such a way that in 90 percent of the network expansion steps the preferential attachment process is utilized, and random attachment process is utilized in the rest of the expansion steps. Consequently, the problem is to find the optimal probability of employing different processes in order to generate graphs most similar to the target network. In other words, we develop a network generation framework in which several network expansion processes are employed based on their corresponding assigned probabilities, in order to generate networks similar to the target graph. Additionally, each network process $c_i$ should also be configured with a parameter set $r_i$, where $r_i$ includes one or more parameters which configures process $c_i$. For example, the preferential attachment process is based on one parameter $m$ which specifies the number of attached links of a new incoming node. Consequently, the process parameter set ($r_i$) should also be optimized along with the probability of process employment ($p_i$) for each process ($c_i$). In our proposed framework, the process probabilities are optimized along with the process parameters in an evolutionary system. 

Table \ref{tab:notations} describes the defined symbols in our solution encoding. Accordingly, having the target network $G_t$ and the set of candidate processes $\{c_i\}$, the problem is to find the optimal corresponding probability values $\{p_i\}$ and parameter sets $\{r_i\}$, so that the emerging process framework can generate graph $G_r$ with minimum $error_m(G_{t} , G_{r})$ value with respect to different topological metrics $m$.

\begin{table*}
	\centering
	\caption{\label{tab:notations}Table of symbols.}
	\begin{tabular}{c| m{0.5\textwidth} }
		Symbol & Description \\ \hline
		$G_{t}$  &  Target network\\ \hline 
		$G_{r}$  &  Synthesized (artificial) network 
		\\ \hline 
		$C=\{c_1,c_2, ..., c_n\}$  &  The set of candidate processes\\ \hline 
		$P=\{p_1,p_2, ..., p_n\}$  &  The set of evolved (estimated) probabilities of candidate processes: $p_i = P(c_i)$ \\ \hline 
		$R=\{r_1,r_2, ..., r_n\}$  &  Optimized parameter sets of the corresponding candidate processes $c_i$ \\ \hline 
		$error_m(G_{1} , G_{2})$  &  The amount of dissimilarity of graphs $G_{1}$ and $G_{2}$ according to the topological property (metric or measurement) m\\ 
	\end{tabular}
\end{table*}

\subsection{Employed Network Processes} \label{sub:networkprocesses}
In this research, we utilize several basic network processes as the building blocks of the proposed framework. The candidate processes contribute to gradually synthesize an artificial graph which is adapted to the topological features of the target network. In each step of our proposed network generation framework, called \textit{NetMix}, a candidate process ($c_i$) is selected based on the process probabilities ($p_i$) along with its optimized parameters set ($r_i$). The selected process is then utilized to attach $n$ new nodes to the rest of the synthesized network for that generation step. As described in section \ref{sec:problemStatement}, a search mechanism is necessary to find the optimal values of $p_i$ and $r_i$. Before describing our optimization method, we first illustrate the candidate processes utilized in this research. It is worth noting that our proposed framework is not limited to the set of employed network processes, and one may add/remove several formation processes to this framework. In order to keep the framework concise, we have chosen only four simple formation processes in our framework. We will show the effectiveness of the employed processes in section \ref{sec:evaluations}.

We prepared four candidate processes, each of which results in emergence of a set of important network features. First, in \textit{``Transitive/Random Attachment'' (TRA)} process, a regular lattice of $n$ nodes with degree $K$ is generated (each node is connected to its $K$ adjacent neighbors) and then, each edge of the lattice is rewired to a randomly-chosen existing node with probability $p_{rewiring}$. Actually, \textit{TRA} process mainly resembles the Watts-Strogatz model \cite{WattsStrogatz} with high clustering and small-world features. Consequently, $K$ and $P_{rewiring}$ are the configuration parameters of the TRA process, which will be optimized in our proposed method. Additionally, the TRA process also represents the random attachment process (inspired by Erd\H{o}s-R\'{e}nyi model \cite{CentralLimit}) since each new edge in this process may be randomly rewired to an existing node with probability $p_{rewiring}$. For example, with a high value of $P_{rewiring}$, this process generates random graphs. 

As the second formation process, we consider \textit{``Preferential Attachment'' (PA)} process which acts similar to Bar\'{a}basi-Albert \cite{BAModel} model. In each generation step of this process, $n$ new nodes are attached to the network based on the degree of the existing nodes. In other words, a new node $v_n$ is attached to an existing node $v_o$ with a probability proportional to the degree of $v_o$ (i.e., ${p(\{v{_n},v_{o}\} \in E)} = {{k_{v_o}} \over {\sum_{v}{k_v}}}$, where $k_x$ is the degree of node $x$). Each new node actually attaches to $m$ existing nodes and thus, PA process is configured with parameter $m$. 

The third defined process is \textit{``Modular Attachment'' (MA)} which is inspired by copying network model \cite{WebAsGraph}. In this process, each new node first randomly selects an existing node $x$, and attaches to each neighbor of $x$ with probability $P_{copying}$. If $x$ has no edges, the new node attaches to $x$ itself. The Modular Attachment process supports modular networks with community structure. 

Finally, the fourth proposed process is  \textit{``Assortative/Disassortative Mixing'' (ADM)} process, in which the degree-assortativity \cite{AssortativityNewman} of the network is increased or decreased. Assortativity is a condition in which the degree of the linked nodes are correlated. In an assortative network, there is a positive correlation between the degrees of two attached nodes, and conversely a disassortative network shows a negative degree correlation between the linked nodes. In ADM process, we simply try to change the assortativity of the network towards the assortativity of the target network. In this regard, a pair of existing nodes are chosen randomly, and their connection (edge) is added or removed in order to make the overall network assortativity closer to the target assortativity. This task is repeated $N_{ADM}$ times, where $N_{ADM}$ is an evolved parameter of this process.
 
  The four considered network processes (TRA, PA, MA, ADM) are capable of generating networks with various network features. For example, PA supports scale-free networks with long-tail degree distribution, TRA supports transitive relationships, small path length, and random (casual) attachments. MA results in modular networks with community structure, and finally ADM supports assortative or disassortative networks. Consequently, a mixture of the four processes is capable of generating networks which are adapted various properties of the target network, while a single process (or network model) may fail to support the mixing features of the target network. Table \ref{tab:generativeparams} shows the process probabilities and configuration parameters which are all optimized in our proposed method described in the next subsection.

\begin{table*}
	\centering
	\caption{\label{tab:generativeparams}The process probabilities notation and the configuration parameters.}
	\begin{tabular}{c| m{0.5\textwidth} }
		Parameter Name & Description \\ \hline
		$n$  &  Number of new nodes (added to the network) in each step of network expansion\\  \hline
		$P_{PA}$  &  Probability of choosing ``Preferential Attachment'' (PA) process in each step of network expansion\\ \hline 
		$m$     & Number of attachments of each new node in PA process\\ \hline
		$P_{TRA}$  &  Probability of choosing ``Transitive/Random Attachment'' (TRA) process in each step of network expansion\\ \hline 
		$K$  &  Number of adjacent neighbors of each node in TRA process\\ \hline
		$P_{rewiring}$  &  The probability of displacing one end of each edge in the new generated regular lattice to a random existing node, in TRA process\\ \hline
		$P_{MA}$  &  Probability of choosing ``Modular Attachment'' (MA)  process in each step of network expansion\\ \hline 
		$P_{copying}$  &  The probability of attaching a new node to each neighbor of a randomly-chosen existing node, in MA process\\ \hline
		$P_{ADM}$  &  Probability of choosing ``Assortative/Disassortative Mixing'' (ADM)  process in each step of network expansion \\ \hline 
		$N_{ADM}$  &  Number of considered pairs in ADM process \\ 
	\end{tabular}
\end{table*}

\subsection{Optimal Mixture of Network Processes} \label{sub:ga}
Our proposed method, named \textit{NetMix}, combines different network processes in order to develop a network model framework which supports a mixture of nontrivial topological features. The configuration of such a model framework should be tuned per target network based on the desired topological characteristics. The configuration setting includes the probability of applying each candidate process, along with the parameters of the processes. Table \ref{tab:generativeparams} illustrated the parameters which should be specified for any target network. As a result, the problem of adaptive network generation is reduced to a search problem for finding the optimal values of those parameters per target network. This problem faces a huge search space because we should find the optimal values for many parameters with a wide range of possible values. Fortunately, meta-heuristic algorithms such as \textit{``Genetic Algorithm''} are known to be effective in optimization of such problems with large search space \cite{GAbook,genetic}. 

Inspired by Darwinian evolutionary theory and the process of natural selection, genetic algorithm (GA) is widely employed in optimization and search problems by simulating bio-inspired operators such as crossover, mutation, and selection. When applying GA for an optimization problem, a population of candidate solutions (called individuals or chromosomes) with an specified set of properties (genes) is evolved. According to the fitness of its solution, each individual is given a score that determines its survival rate in subsequent generations. In each generation of GA, ``parent'' solutions are also selected from the existing individuals according to their fitness score, for breeding using crossover and mutation operators.

In our proposed method, we search the optimized values for the parameters described in Table \ref{tab:generativeparams} and thus, we represent a chromosome by a vector of those parameters. We defined appropriate crossover and mutation operators along with a fitness function. In each evolution generation, we employ ``tournament selection'' \cite{genetic} to choose the parents and ``uniform crossover''\cite{GAbook,genetic} to generate new child individuals. For the mutation operator, we change some genes of an individual to random values. More details about the configuration of the implemented genetic algorithm is illustrated in Section \ref{sub:experimentdetails}. In order to calculate the fitness of an individual, we generate a graph using the configuration parameters encoded in that individual chromosome, and then we compute the similarity of the target network to the generated graph. We utilize \textit{NetDistance} \cite{netdistance} as the fitness function for calculating the topological similarity of two complex networks. NetDistance \cite{netdistance} defines the amount of dissimilarity of two complex networks equal to the weighted Manhattan distance of some of their topological features including degree distribution, average clustering coefficient, transitivity, assortativity, and modularity: $netdistance(g_{1},g_{2}) = \sum_{i=1}^{n} w_{i}| m_{i}(g_1)-m_{i}(g_2) |$, where $m_{i}(g)$ is the $i^{th}$ considered property (metric) of the graph $g$ and $w_i$ is its corresponding learned weight (trained in a machine learning algorithm). NetDistance is a size-independent dissimilarity metric (distance function), which is capable of comparing the overall topological features of complex networks\cite{netdistance,modelfit}. It is worth noting that we can extend NetDistance with other properties that contribute in the distance function. Actually, the distance function can be customized according to the employed candidate processes in the proposed method. We will show in the evaluation results (Section \ref{sec:evaluations}) that the simple utilized distance function (NetDistance) is an effective fitness measure for synthesizing networks similar to the target graphs.

\section{Evaluations} \label{sec:evaluations}

\subsection{Baseline Methods} \label{sub:baselines}
In order to investigate effectiveness of the proposed method, we compare it with several baselines. As the first category of baselines, six network models are considered in our experiments: Barab\'{a}si-Albert (BA) model \cite{BAModel}, Erd\H{o}s-R\'{e}nyi (ER) \cite{CentralLimit}, Forest Fire (FF) \cite{graphsOverTime}, Kronecker graphs model (KG) \cite{kronecker}, Random power-law (RP) \cite{RandomPowerLaw}, and Watts-Strogatz (WS) model of Small-World networks \cite{WattsStrogatz}. These models are compared with our proposed method according to their effectiveness in generating artificial networks similar to the target network. This comparison is performed based on different topological measurements of complex networks such as degree distribution and community structure. The six models are chosen mainly because they are frequently used in network generation applications, and they also cover a wide range of network structures, such as scale-free graphs and small-world networks.

In addition to the six mentioned network models, we also consider ABNG method \cite{arora2017action} as another baseline method in our evaluations. This is because ABNG is a successful and more recent attempt which follows an adaptive approach of network generation. As explained in Section \ref{sec:relatedworks}, ABNG uses simulated annealing in order to optimally orchestrate simple actions of network formation. ABNG shows remarkable results for generating graphs similar to various artificial and real networks.

\subsection{Observed Network Properties} \label{sub:networkproperties}
Our proposed method can replicate different network properties of the target networks. In order to evaluate this capability, we compare the synthesized graphs with corresponding target networks according to different network properties. We define the $error_m$ in replicating feature (metric) $m$ as the amount of dissimilarity of the target network to the adapted graph (i.e., the corresponding network synthesized by our proposed method) according to the metric $m$. First, we consider several global graph properties including average clustering coefficient \cite{WattsStrogatz}, transitivity \cite{surveyOfMeasurements}, modularity \cite{ModularityNewman}, and assortativity (degree correlation) \cite{AssortativityNewman}. For each of the mentioned global metrics $m$, $error_m$ is defined as the absolute difference of the measure $m$ between the target and the synthesized network. 

Additionally, in order to replicate evaluation results of our main baseline (ABNG \cite{arora2017action}), we considered several node properties including node degree, local-clustering, closeness centrality, 	 centrality, eigenvector centrality, and PageRank centrality. Then, we quantify the two-sample Kolmogorov-Smirnov (KS) test of the distribution of the mentioned node properties between the synthesized and target graphs. Moreover, we utilized the DDQC method \cite{ddqc} as another measure for comparison of two degree distributions (in addition to the KS-test of the degree distributions). DDQC (Degree Distribution Quantification and Comparison) considers two networks with probably different sizes, normalizes their degree distributions, and then compares the distributions. It is shown that DDQC is more effective than KS-test in comparing the degree distribution of complex networks \cite{ddqc}. 

\subsection{Target Networks Dataset} \label{sub:targetnetworks}
We consider different artificial and real networks as the target graphs in order to evaluate the effectiveness of the proposed method in reproducing the properties of the target graphs. As artificial target graphs, we consider many networks generated by the six mentioned network models described in Section \ref{sub:baselines} (BA\cite{BAModel}, ER\cite{CentralLimit}, WS\cite{WattsStrogatz}, FF\cite{graphsOverTime}, KG\cite{kronecker}, and RP\cite{RandomPowerLaw}). 

In addition to the artificial graphs, we consider 17 real-world networks as the target networks in our evaluations. The set of selected real networks contains diverse networks from different network types. Moreover, we had access to the evaluation results of the baseline method on the selected real networks and therefore, we were able to compare the evaluation results of our proposed method with the baseline (the baseline also includes evaluating three brain networks \cite{bellec2017neuro} but we had no access to those networks nor the evaluation results and therefore we excluded them). The dataset of considered real networks includes: The network of co-appearing characters in the novel Les Mis\'erables \cite{CoappearanceNetwork}, a network of US political books sold Amazon.com \cite{PolBooksAmazon}, the network of common adjectives and nouns in the novel David Copperfield by Charles Dickens (adjacent words network) \cite{wordadj}, a network of American football games \cite{GirvanNewman}, a network of collaborations between the Jazz musicians \cite{JazzCollaborations}, a network of social relations between dolphins living in Doubtful Sound, New Zealand \cite{DolphinsSocialNet}, two protein networks of C-alpha atoms (1php and 1qop) \cite{phpqop},  a yeast protein interaction network \cite{YeastProtein}, networks of Biogrid FRET, Far Western and Dosage Lethality from a general repository for interaction datasets \cite{biogrid}, a network which represents flights between US Airports \cite{USAirport2007}, Norwegian boards network which represents relationships between board members of norwegian public companies in August 2011 \cite{NorwegianBoardsNet}, a human protein interaction network \cite{HumanProtein}, a social network of students at University of California, Irvine \cite{SocialNetwork}, and a power-grid network from the United States \cite{WattsStrogatz}.

\subsection{Evaluation Results} \label{sub:results}
\begin{figure*}
	\subfigure[]
	{    
		\includegraphics[width=0.3\textwidth]{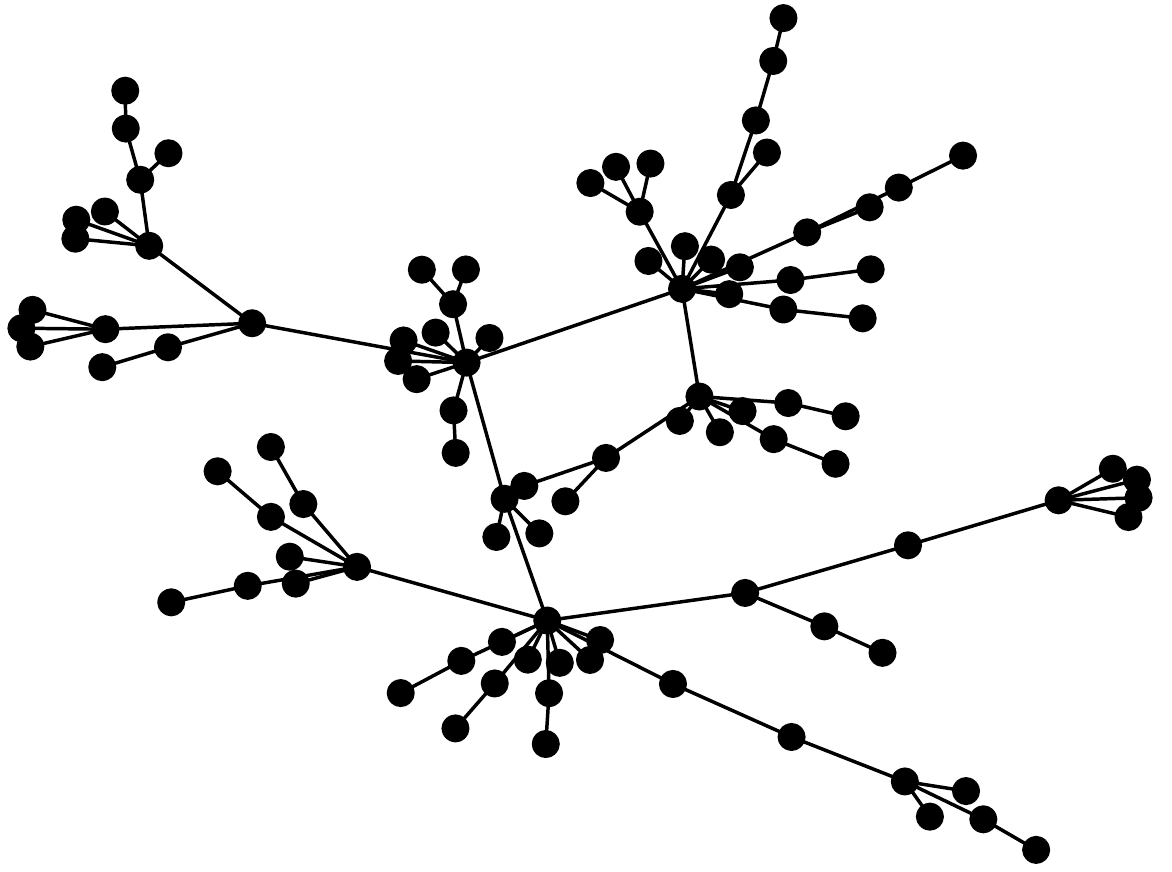}
		\label{fig:evalBASampleGen}
	}
	\subfigure[]
	{    
		\includegraphics[width=0.3\textwidth]{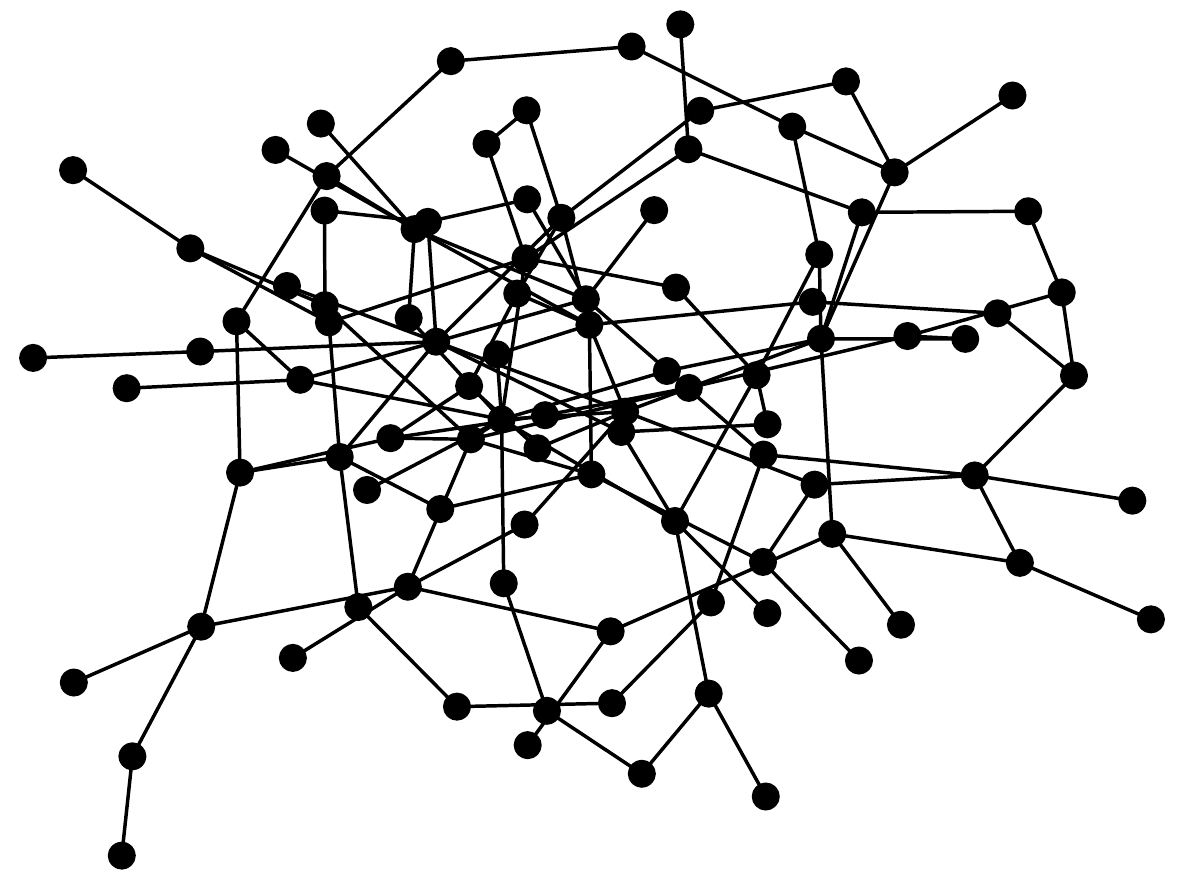}
		\label{fig:evalERSampleGen}
	}
	\subfigure[]
	{    
		\includegraphics[width=0.3\textwidth]{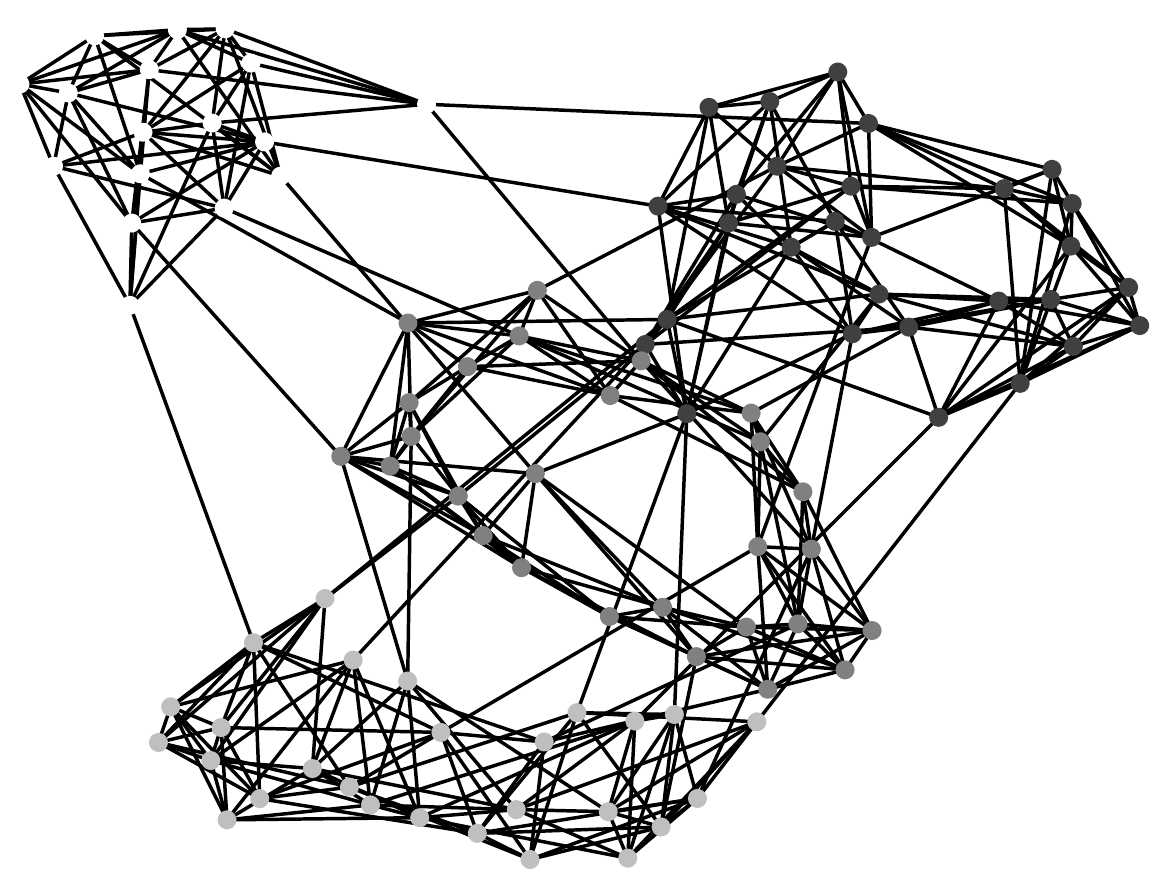}
		\label{fig:evalWSSampleGen}
	}
	\caption{Sample graphs generated by our proposed method which are adapted to different target networks. \subref{fig:evalBASampleGen} The target graph is a scale-free graph synthesized by Barab\'{a}si-Albert (BA) model \cite{BAModel} \subref{fig:evalERSampleGen} The target network is a random graph generated by Erd\H{o}s-R\'{e}nyi (ER) \cite{CentralLimit} \subref{fig:evalWSSampleGen} The target network is a small-world graph generated by Watts-Strogatz (WS) model \cite{WattsStrogatz}.}
	\label{fig:sampleGenGraphs}
\end{figure*}

In contrast to classical manual-designed network models which only support a limited kind of networks, such as Barab\'{a}si-Albert (BA), our proposed method is an adaptive and automated approach of network generation. However, we begin the evaluations by investigating the ability of our proposed method to reproduce network features of artificial graphs which are generated by classical manual-designed network models. This evaluation is important because of the historical significance of the network models and the specific network features that they model \cite{arora2017action}. First of all, we simply show sample graphs generated by our proposed framework (NetMix) which are illustrated in Figure \ref{fig:sampleGenGraphs}. When the target graph is a scale-free network synthesized by Barab\'{a}si-Albert (BA) model \cite{BAModel}, NetMix is capable of generating similar graphs with characteristic hierarchical structure of scale-free networks (Figure \ref{fig:evalBASampleGen}). If the target graph is a random graph, NetMix also generates a graph with a random-like topology (Figure \ref{fig:evalERSampleGen}). Finally, if the target graph is a small-world graph generated by Watts-Strogatz (WS) model \cite{WattsStrogatz}, NetMix synthesizes graphs with high clustering and small path-lengths (Figure \ref{fig:evalWSSampleGen}). As the figures show, sample synthesized graphs are successfully adapted towards the target networks. 

\begin{figure*}
	\subfigure[]
	{    
		\includegraphics[width=0.4\textwidth]{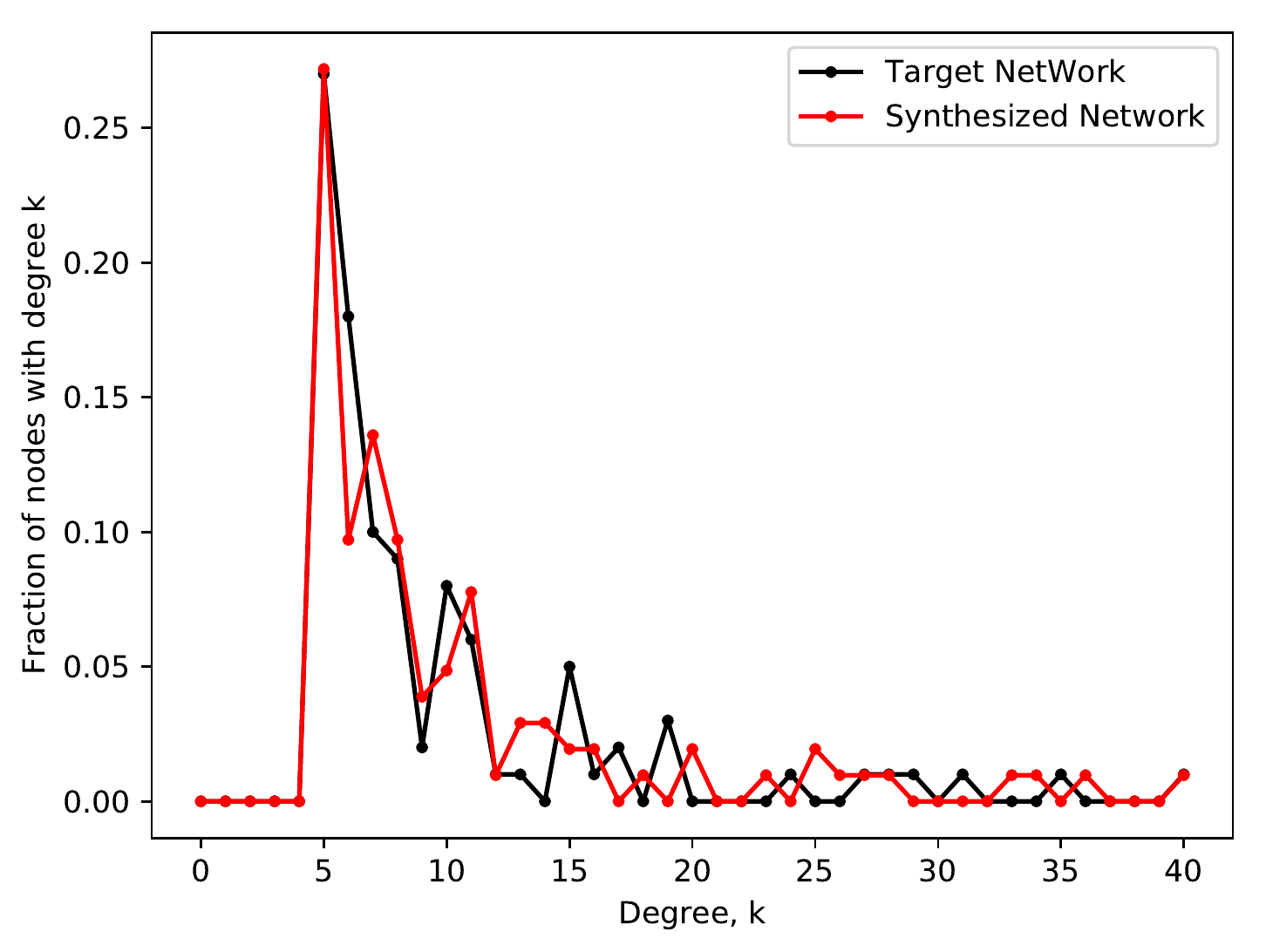}
		\label{fig:degDistBA}
	}
	\subfigure[]
	{    
		\includegraphics[width=0.4\textwidth]{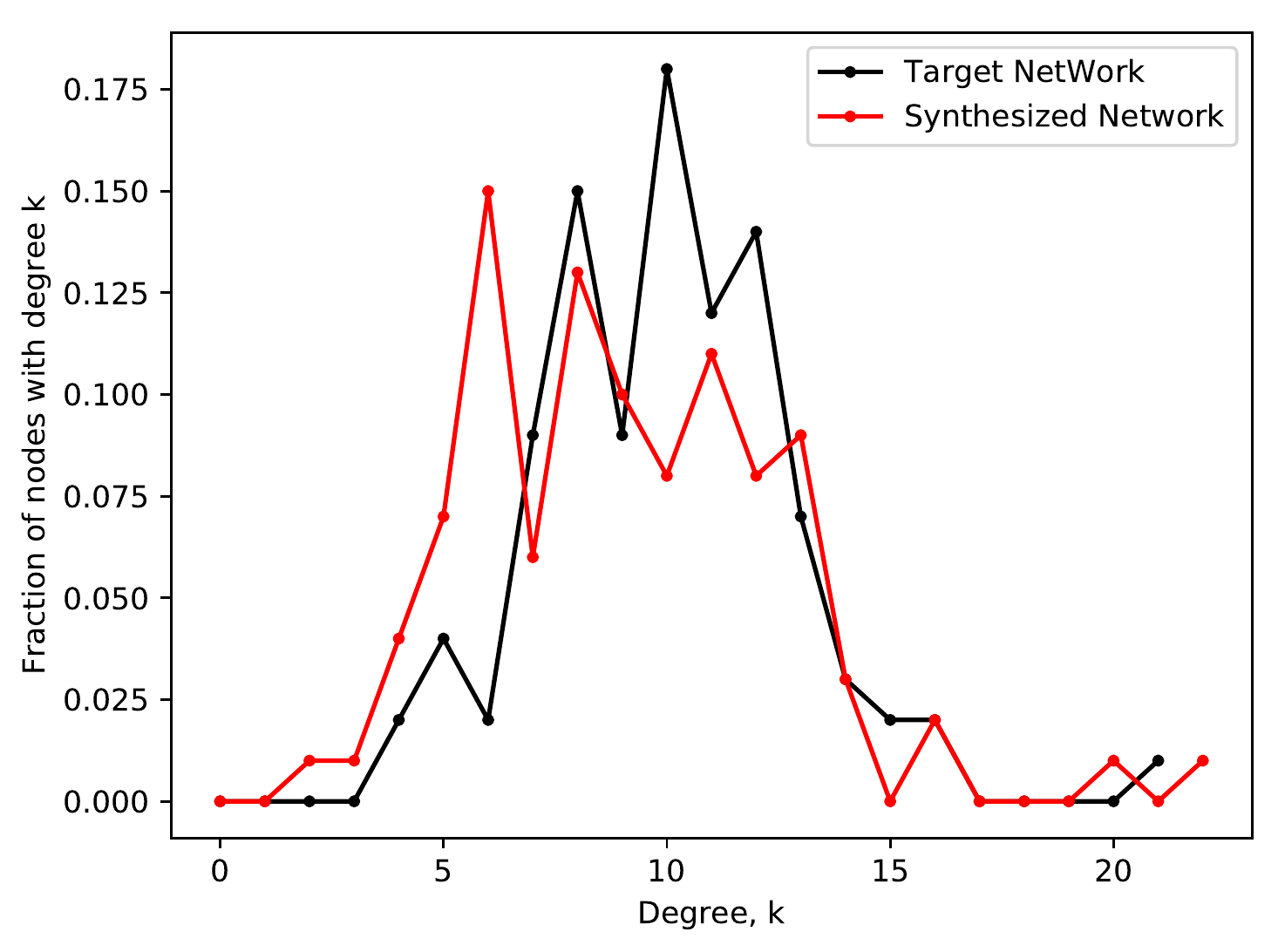}
		\label{fig:degDistER}
	}
	\subfigure[]
	{    
		\includegraphics[width=0.4\textwidth]{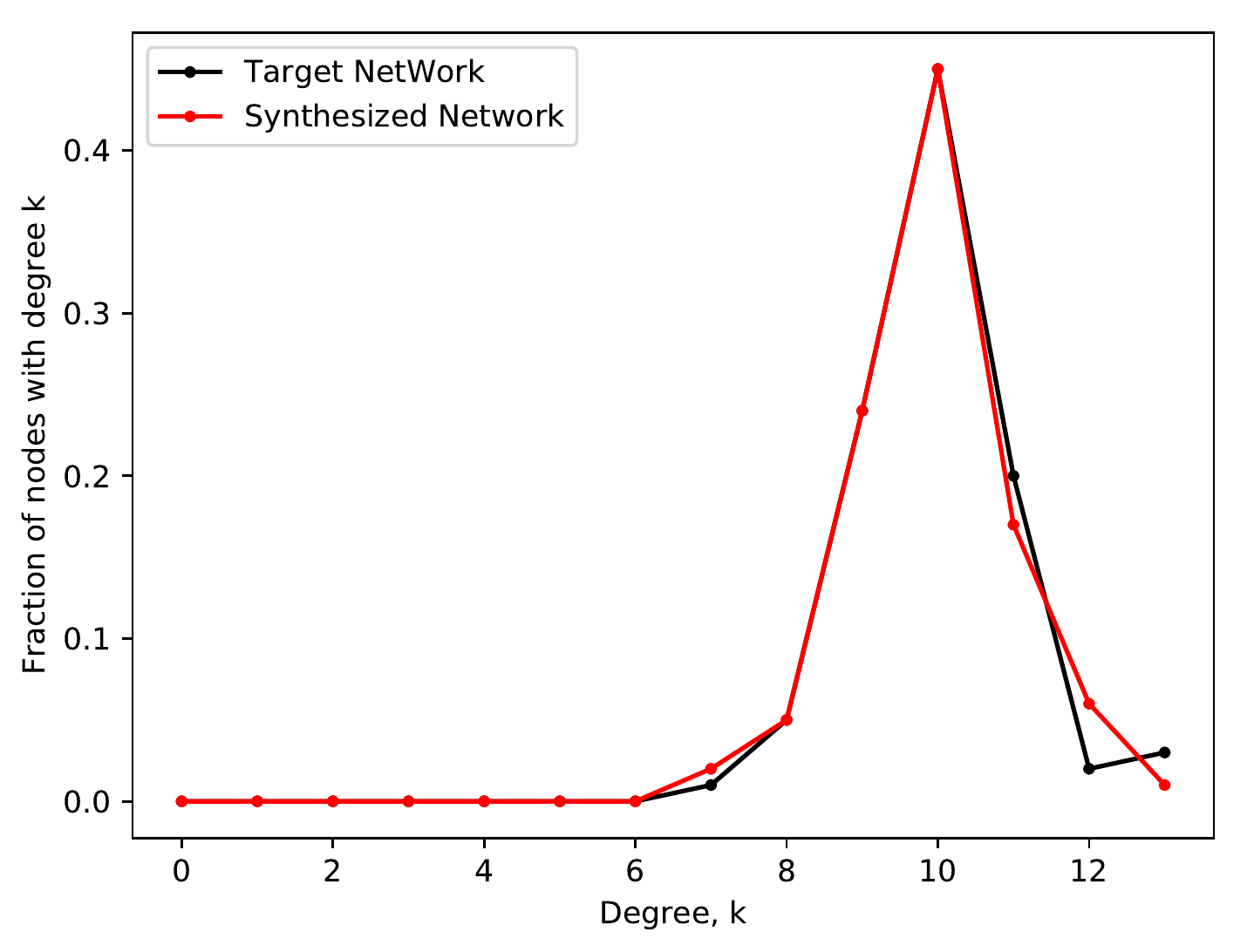}
		\label{fig:degDistWS}
	}
	\subfigure[]
	{    
		\includegraphics[width=0.4\textwidth]{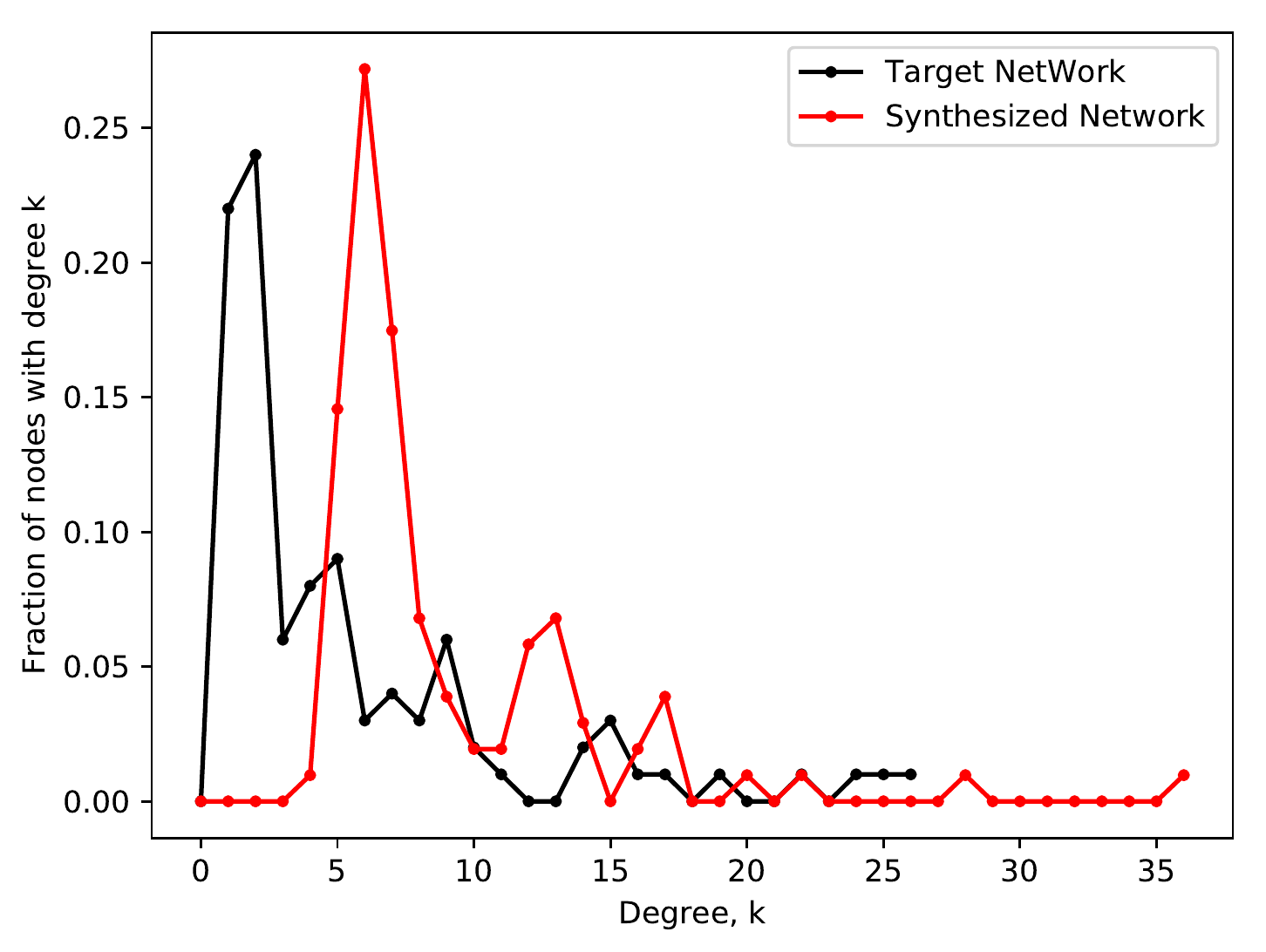}
		\label{fig:degDistFF}
	}
	\subfigure[]
	{    
		\includegraphics[width=0.4\textwidth]{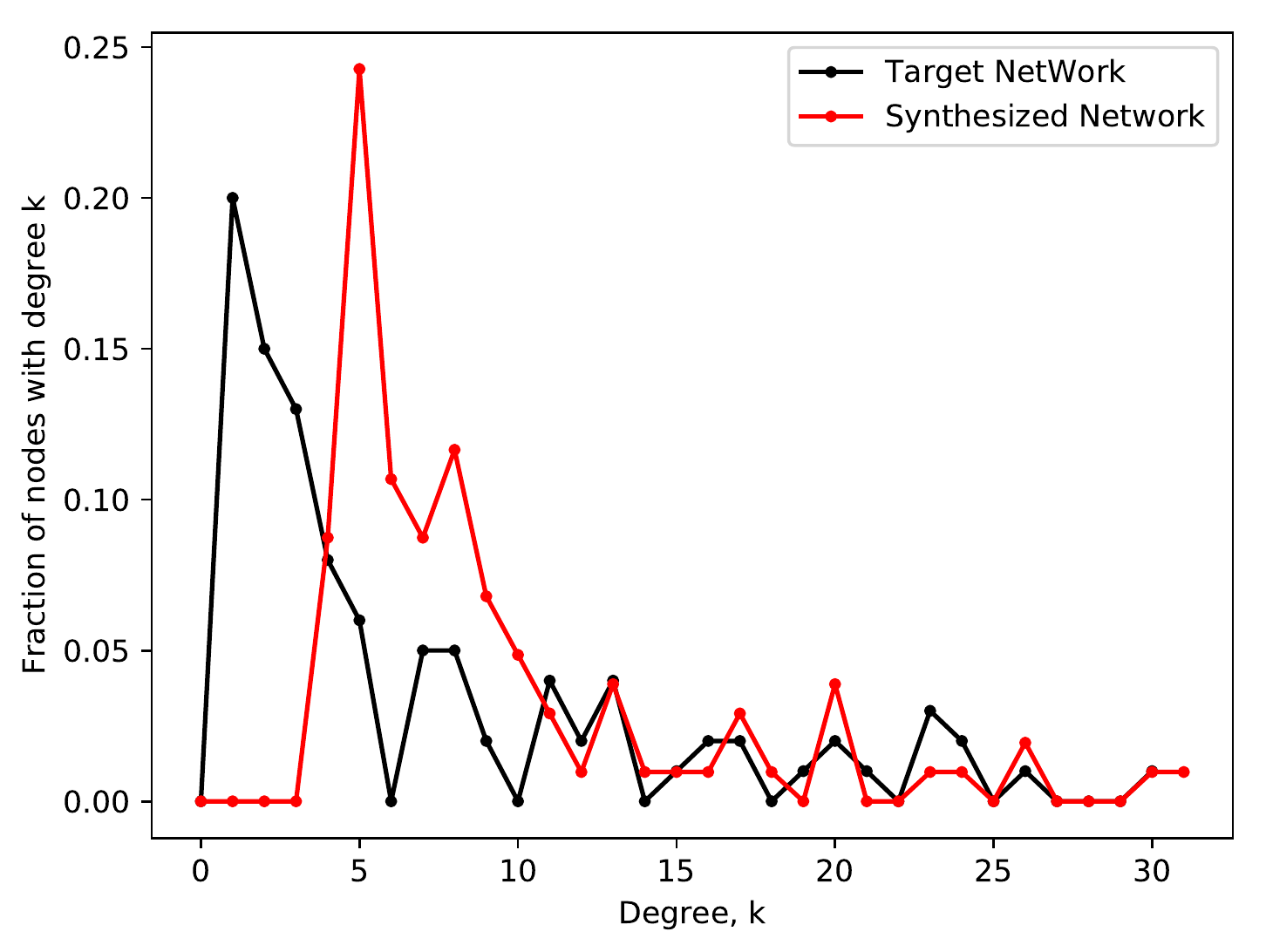}
		\label{fig:degDistRP}
	}
	\subfigure[]
	{    
		\includegraphics[width=0.4\textwidth]{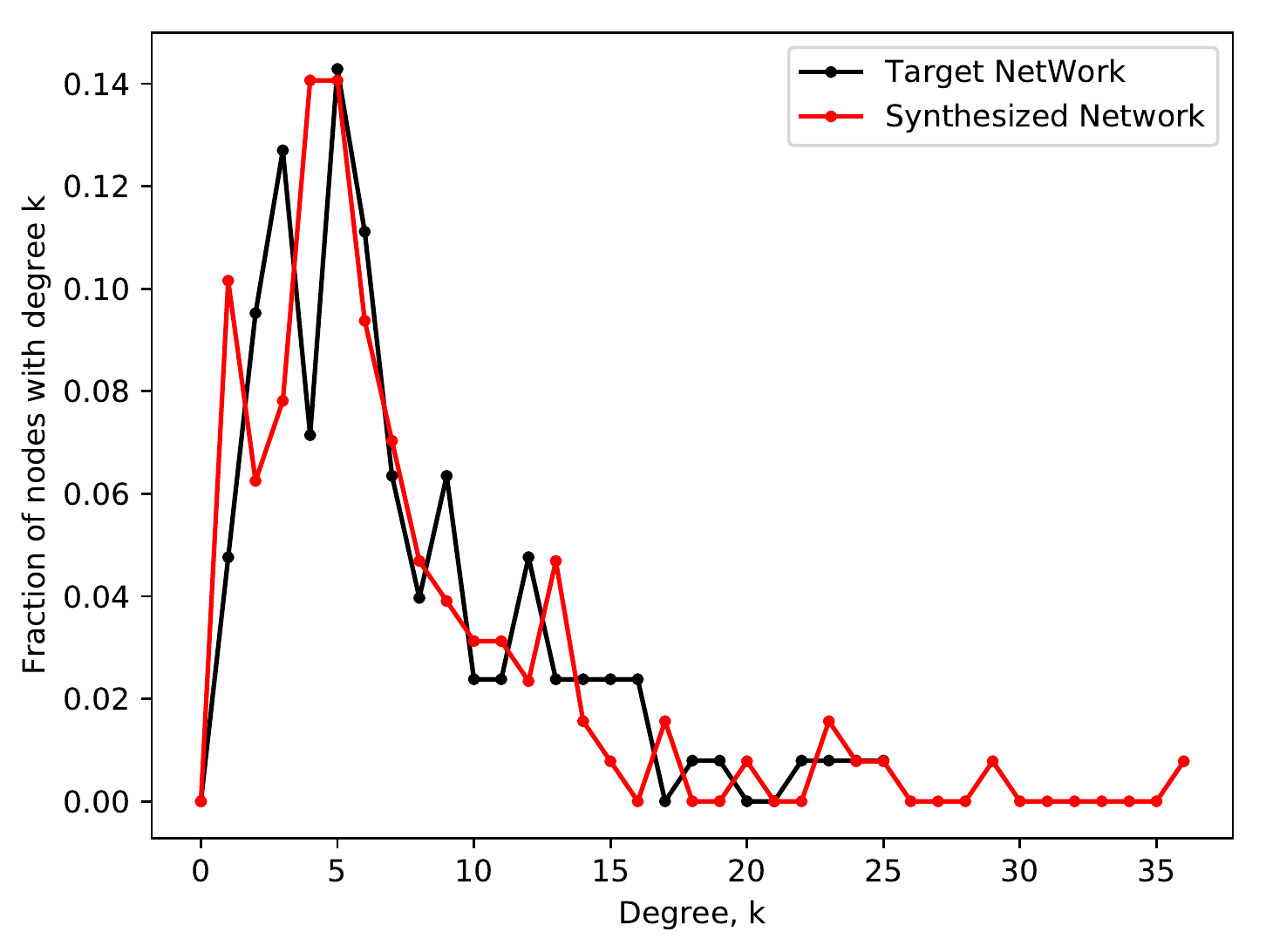}
		\label{fig:degDistKG}
	}
	\caption{The degree distribution of networks synthesized by NetMix compared to their corresponding target graphs. The target networks are artificial graphs generated by various network models: \subref{fig:degDistBA} Barab\'{a}si-Albert \cite{BAModel} \subref{fig:degDistER} Erd\H{o}s-R\'{e}nyi \cite{CentralLimit} \subref{fig:degDistWS} Watts-Strogatz \cite{WattsStrogatz} \subref{fig:degDistFF} Forest Fire \cite{graphsOverTime} \subref{fig:degDistRP} Random power-law \cite{RandomPowerLaw} \subref{fig:degDistKG} Kronecker graphs \cite{kronecker}
}
	\label{fig:degDist}
\end{figure*}

In order to further investigate the properties of sample generated networks, Figure \ref{fig:degDist} compares the target artificial graphs with the networks synthesized by NetMix framework according to their degree distribution. In this experiment, six sample graphs are considered which are generated by the six network models described in the baseline methods (refer to Section \ref{sub:baselines}). Except the networks generated by WS \cite{WattsStrogatz} and ER \cite{CentralLimit} models, other models result in networks with long-tail degree distribution. As the figure shows, the networks synthesized by NetMix are able to mimic the degree distribution of the target networks, either for the long-tail distribution of BA \cite{BAModel}, KG \cite{kronecker}, RP \cite{RandomPowerLaw}, and FF \cite{graphsOverTime} networks, and for the semi-normal distribution of WS \cite{WattsStrogatz} and ER \cite{CentralLimit} networks. 

\begin{figure*}
	\includegraphics[width=0.6\textwidth]{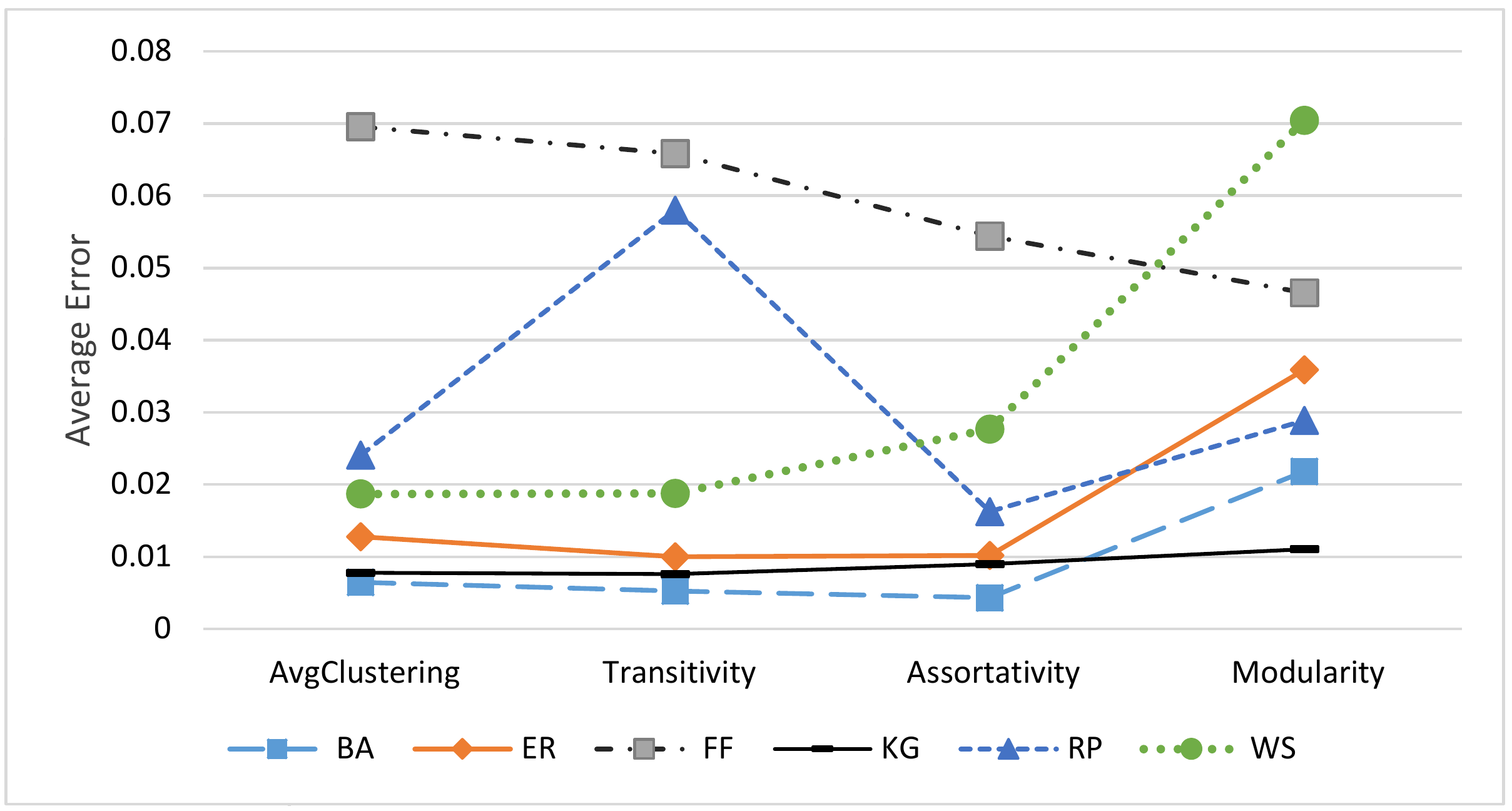}
	\caption{Average error of different networks synthesized by NetMix in various properties of artificial networks.}
	\label{fig:avgErrorArt}
\end{figure*}

The average error of the proposed method ($error_m$) for different global metrics (m) of the artificial networks is also illustrated in Figure \ref{fig:avgErrorArt}. In this experiment, the average error of NetMix is computed when the target networks are artificial graphs generated by the six network models. Figure \ref{fig:avgErrorArt} summarizes the average $error_m$ for 60 different artificial target networks (10 target networks per network model). As the figure shows, $error_m$ is less than 0.07 for all of the considered global properties. Additionally, for most of the compared properties, $error_m<0.03$. 

\begin{figure*}
	\includegraphics[width=0.5\textwidth]{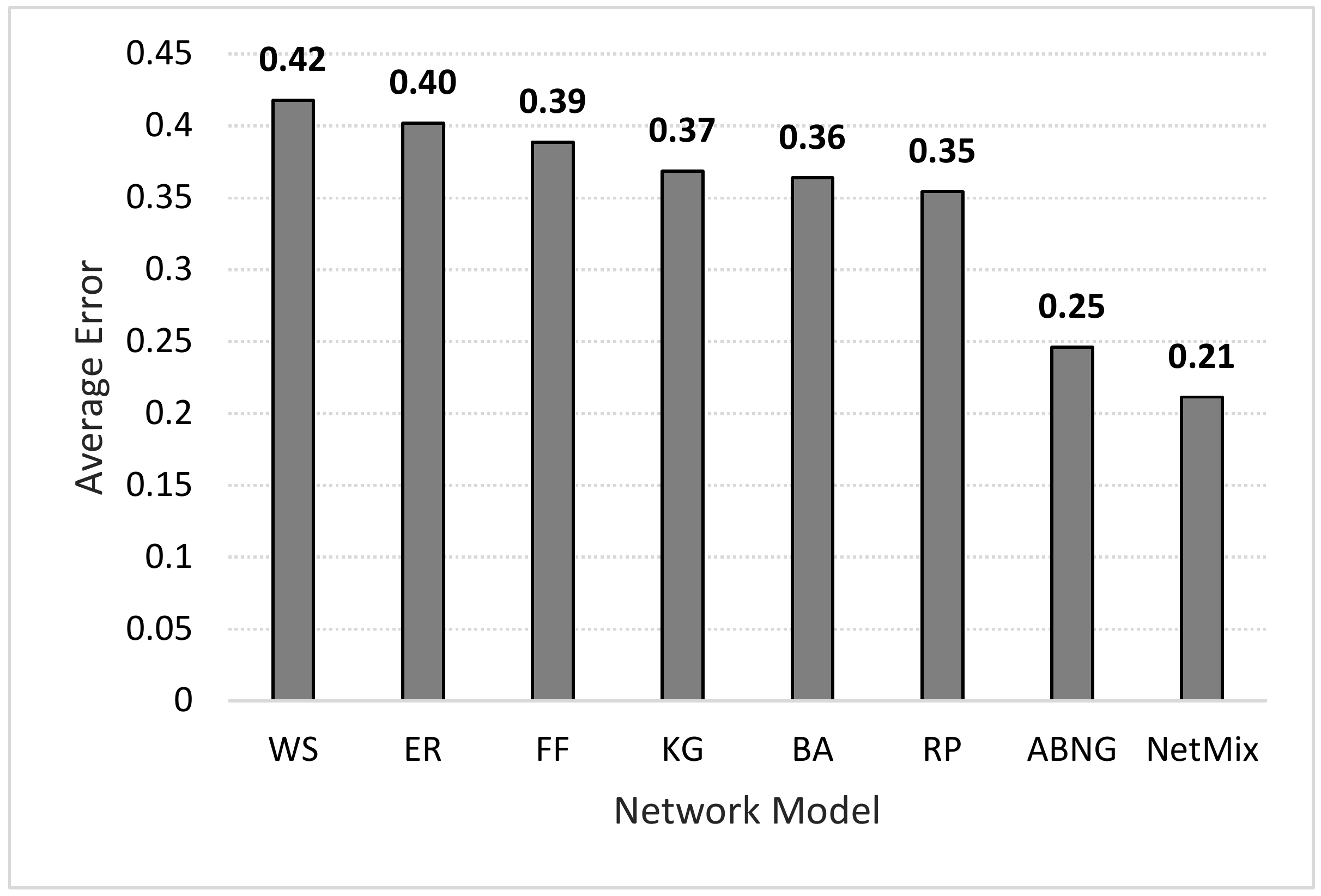}
	\caption{Average error of different methods for various properties in the real-world networks dataset.}
	\label{fig:avgErrorReal}
\end{figure*}

Complex real-world networks show various features which should be modeled by network generators. Therefore, we evaluate our proposed network generation framework by investigating its ability to reproduce important properties of real-networks. Table \ref{tab:propDistancesReal} shows $error_m$ of the proposed method (NetMix) and the baselines for different network measurements $m$ in the real-networks dataset described in Section \ref{sub:targetnetworks}. The presented error values are averaged over all the graphs in the real-networks dataset. In the case of the classical baseline models (i.e., ER \cite{CentralLimit}, WS \cite{WattsStrogatz}, BA \cite{BAModel}, KG \cite{kronecker}, RP \cite{RandomPowerLaw}, and FF \cite{graphsOverTime}) the best parameter sets are first estimated based on the ModelFit method \cite{modelfit}. In other words, the best parameters are first tuned for the target graph, and then the tuned parameters are utilized to generate similar graphs. As Table \ref{tab:propDistancesReal} shows, our proposed method (NetMix) outperforms baselines according to $error_m$ for most of the network properties, and only for three out of eleven network metrics, NetMix is ranked the second best model. After NetMix, ABNG shows the least error for four properties, and Random power-law model shows the best result for only one property. Moreover, Figure \ref{fig:avgErrorReal} summarizes the average $error_m$ of the considered methods for different network properties. As the figure shows, NetMix results in the least average error in comparison to other baselines. 

\begin{figure*}
	\subfigure[Biogrid Dosage Lethality]
	{    
		\includegraphics[width=0.37\textwidth]{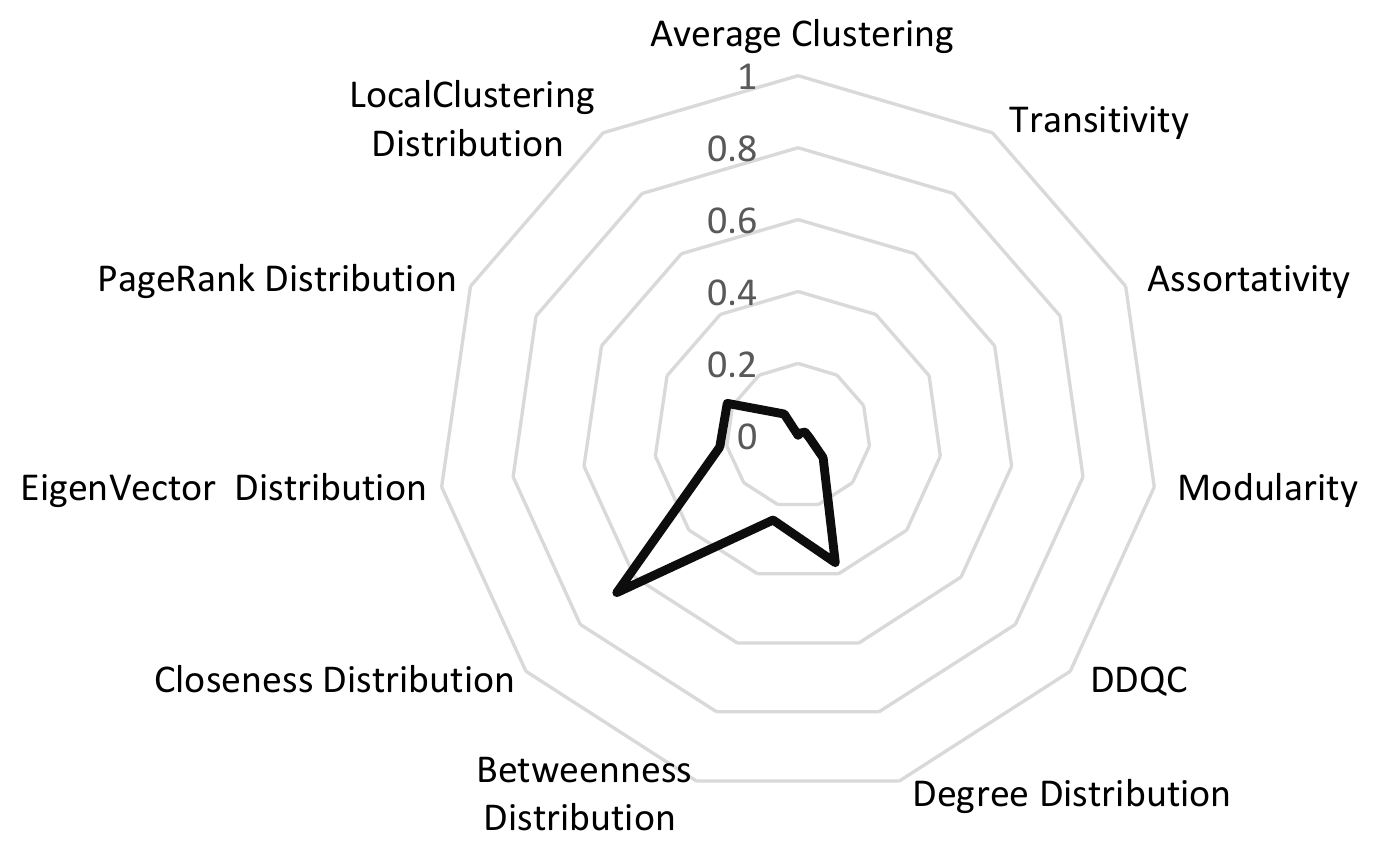}
		\label{fig:radar1}
	}
	\subfigure[American Football]
	{    
		\includegraphics[width=0.4\textwidth]{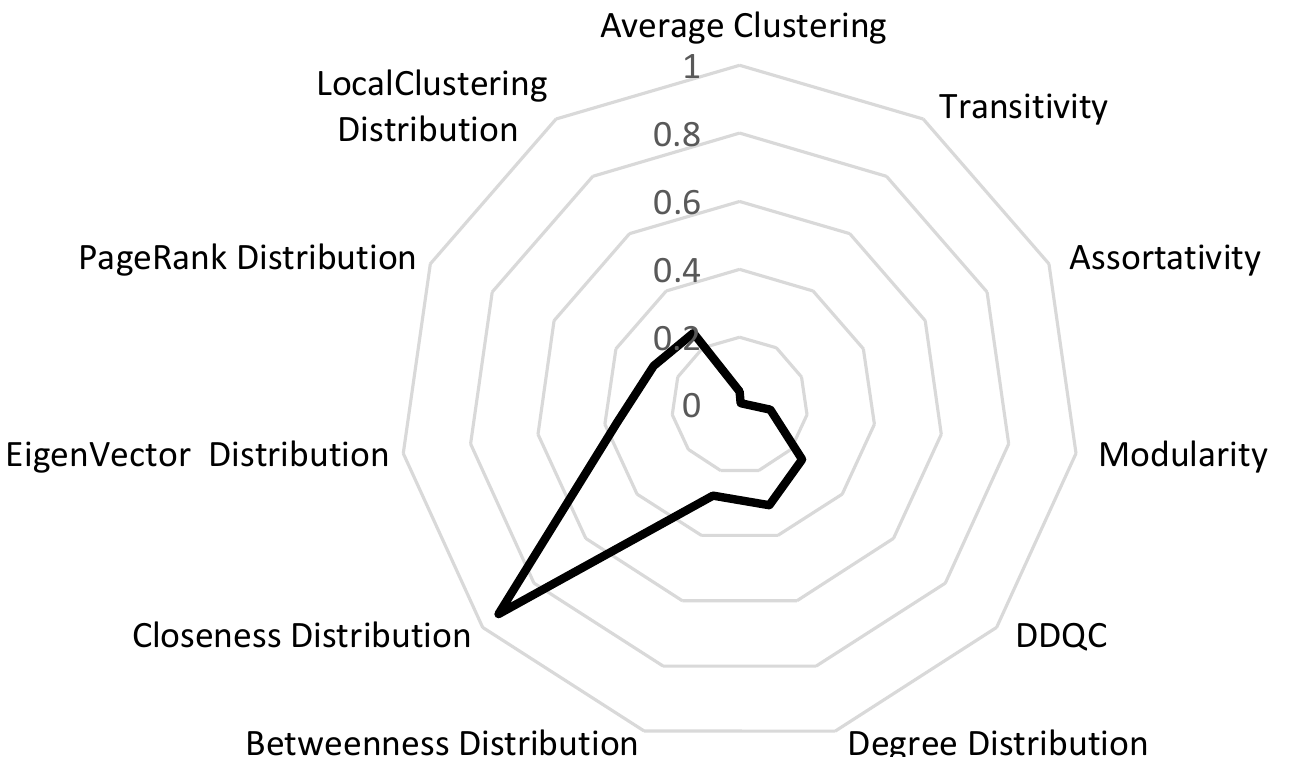}
		\label{fig:radar2}
	}
	\subfigure[Biogrid FRET]
	{    
		\includegraphics[width=0.4\textwidth]{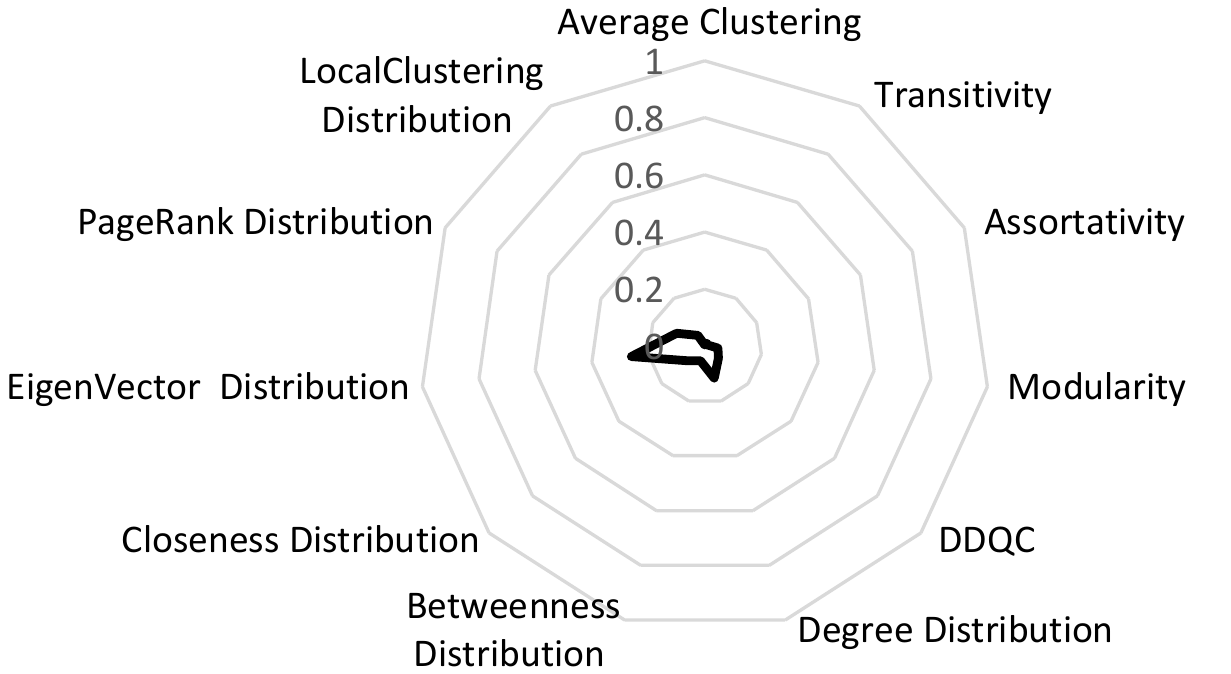}
		\label{fig:radar3}
	}
	\subfigure[Human Protein]
	{    
		\includegraphics[width=0.4\textwidth]{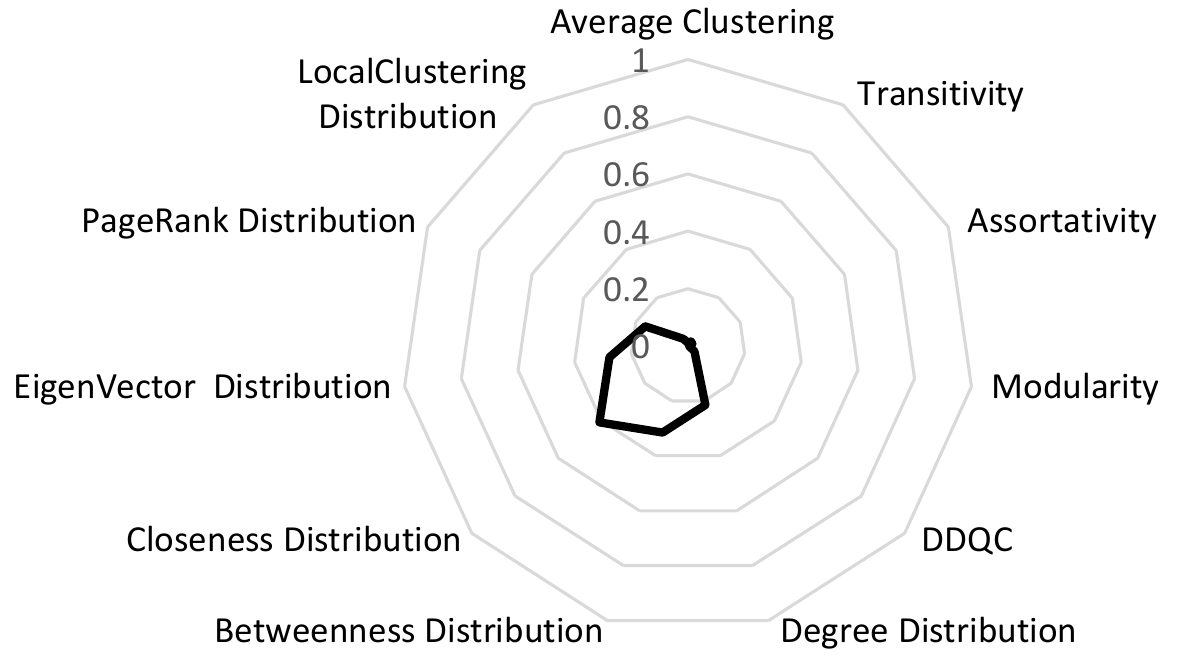}
		\label{fig:radar4}
	}
	\subfigure[UCI Social Network]
	{    
		\includegraphics[width=0.4\textwidth]{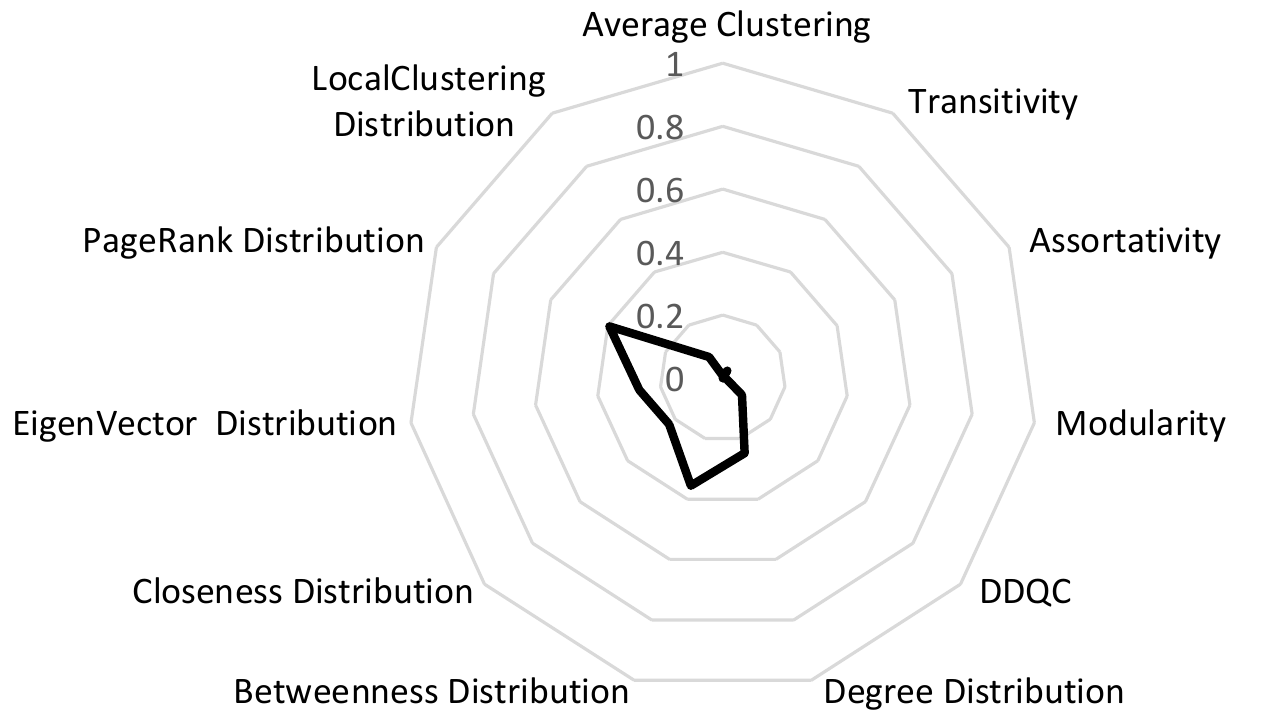}
		\label{fig:radar5}
	}
	\subfigure[Word Adjacencies]
	{    
		\includegraphics[width=0.4\textwidth]{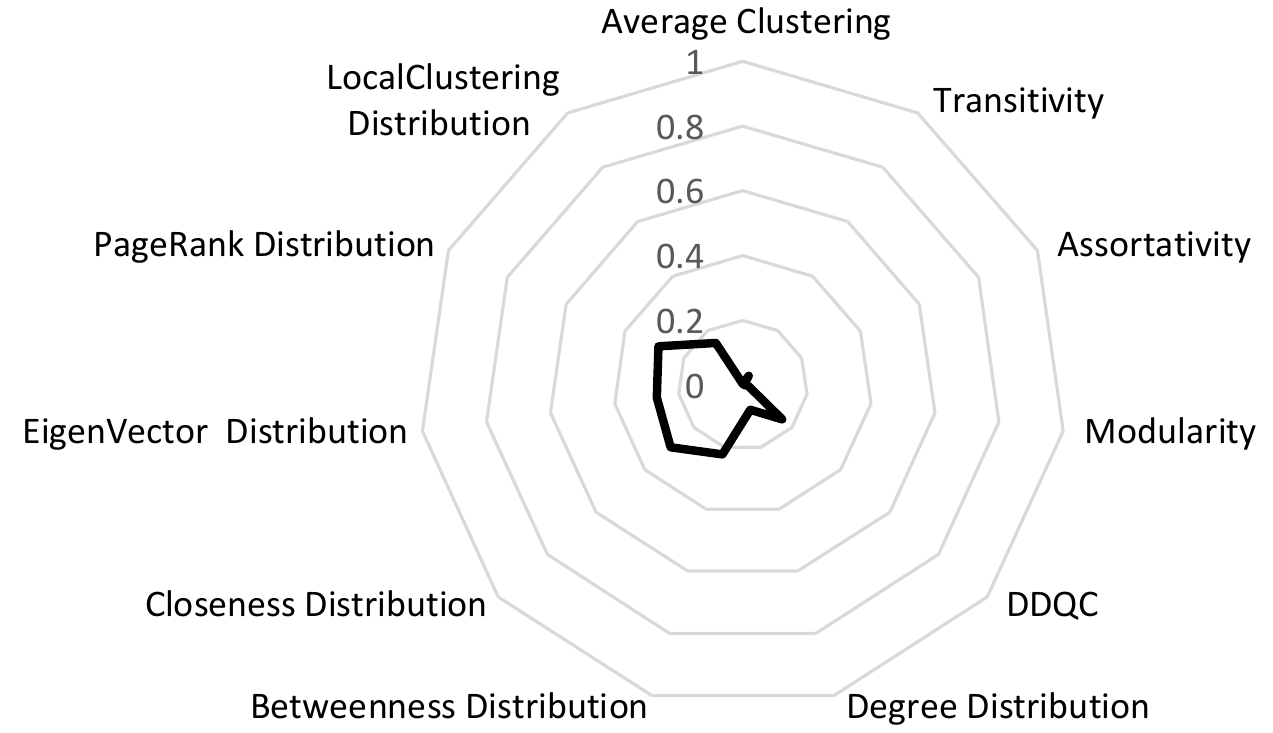}
		\label{fig:radar6}
	}
	\caption{Error of the proposed method in reproducing different network properties for six real-world networks}
	\label{fig:radars}
\end{figure*}

In a more detailed chart, Figure \ref{fig:radars} shows the $error_m$ of NetMix for different properties of several real-world networks. As the radar charts show, the effectiveness of the proposed method is different in various target networks. For example, the proposed method has generated graphs which successfully replicate various properties of Biogrid FRET network, but it has been less successful in American Football network. 

\begin{table*}
	\caption{The average error of different methods based on various network properties (metrics) for the real-world networks dataset.}
	\label{tab:propDistancesReal}
\footnotesize
\begin{tabular}{|l|c|c|c|c|c|c|c|c|}

\hline
\textit{\textbf{Property/Model}} & \textbf{NetMix} & \textbf{ABNG} & \textbf{BA}   & \textbf{ER}   & \textbf{FF}   & \textbf{KG}   & \textbf{RP}   & \textbf{WS}   \\ \hline
\textbf{AvgClustering}           & \cellcolor{lightgray!50}\textbf{0.06}   & 0.11          & 0.19          & 0.24          & 0.18          & 0.22          & 0.20          & 0.15          \\ \hline
\textbf{Transitivity}            & \cellcolor{lightgray!50}\textbf{0.04}   & 0.09          & 0.16          & 0.19          & 0.16          & 0.17          & 0.15          & 0.11          \\ \hline
\textbf{Assortativity}           & \cellcolor{lightgray!50}\textbf{0.03}   & 0.12          & 0.17          & 0.17          & 0.26          & 0.18          & 0.18          & 0.17          \\ \hline
\textbf{Modularity}              & \cellcolor{lightgray!50}\textbf{0.04}   & 0.15          & 0.20          & 0.20          & 0.27          & 0.17          & 0.15          & 0.14          \\ \hline
\textbf{DDQC}                    & \cellcolor{lightgray!50}\textbf{0.31}   & 0.44          & 0.63          & 0.72          & 0.64          & 0.68          & 0.79          & 1.03          \\ \hline
\textbf{Degree Distribution}     & \cellcolor{lightgray!50}\textbf{0.21}   & \cellcolor{lightgray!50}\textbf{0.21} & 0.31          & 0.45          & 0.41          & 0.35          & 0.30          & 0.38          \\ \hline
\textbf{Betweenness Distribution}             & 0.28            & \cellcolor{lightgray!50}\textbf{0.18} & 0.43          & 0.41          & 0.34          & 0.33          & 0.38          & 0.44          \\ \hline
\textbf{Closeness Distribution}               & \cellcolor{lightgray!50}\textbf{0.54}   & \cellcolor{lightgray!50}\textbf{0.54} & 0.66          & 0.65          & 0.85          & 0.69          & 0.67          & 0.72          \\ \hline
\textbf{EigenVector Distribution}             & 0.35   & 0.38          & 0.47          & 0.57          & 0.47          & 0.40          & \cellcolor{lightgray!50}\textbf{0.31}          & 0.61          \\ \hline
\textbf{PageRank Distribution}                & 0.23            & \cellcolor{lightgray!50}\textbf{0.20} & 0.39          & 0.31          & 0.27          & 0.36          & 0.32          & 0.39          \\ \hline
\textbf{LocalClustering Distribution}        & \cellcolor{lightgray!50}\textbf{0.22}   & 0.28          & 0.41          & 0.52          & 0.41          & 0.49          & 0.45          & 0.46          \\ \hline
\end{tabular}
\end{table*}

\begin{figure*}
	\includegraphics[width=0.5\textwidth]{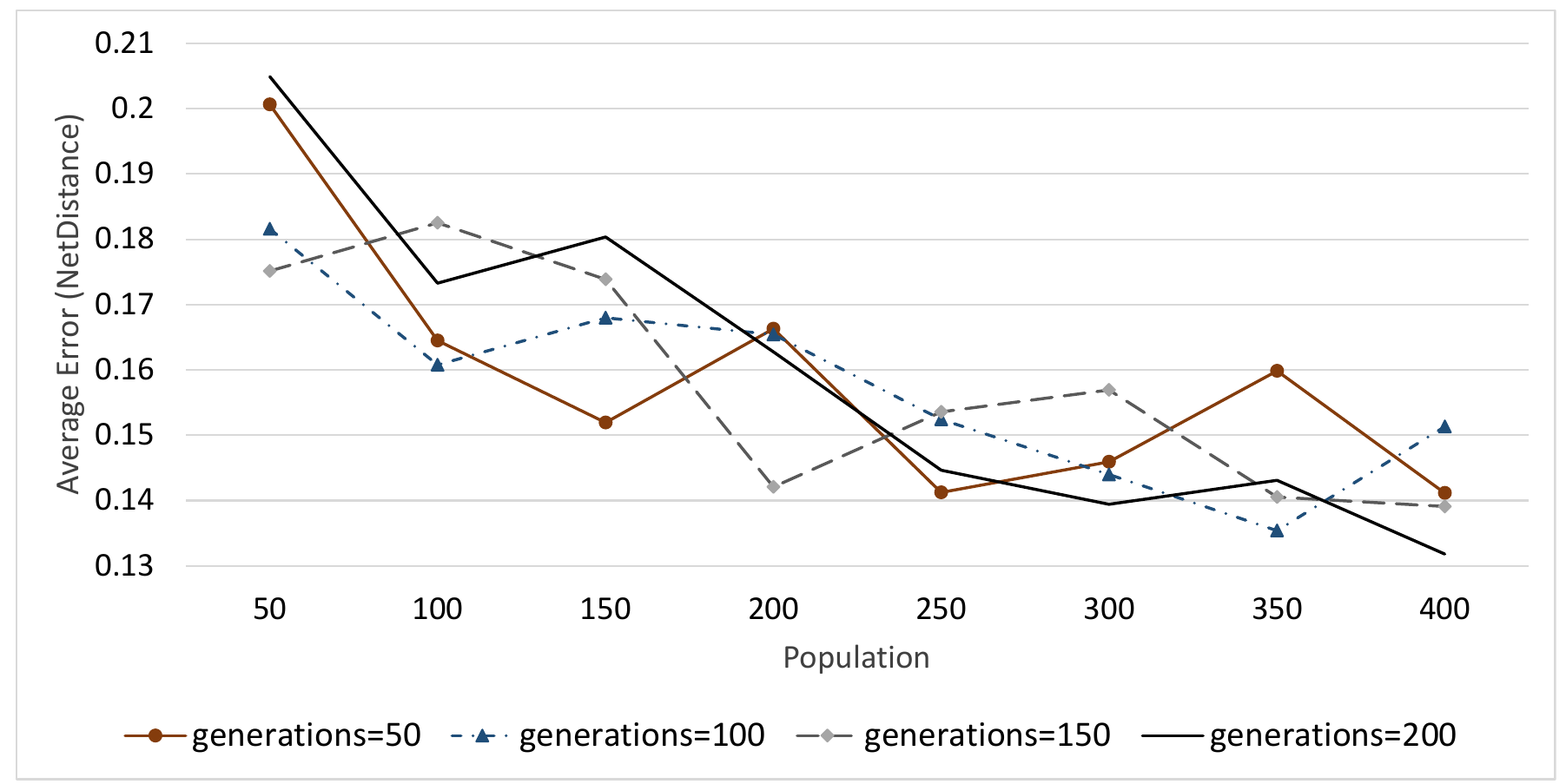}
	\caption{Sensitivity analysis of the proposed genetic algorithm to the population size and the number of generations.}
	\label{fig:sensitivity}
\end{figure*}

Finally, it is also worth evaluating the sensitivity of NetMix to the utilized configuration variables of the proposed genetic algorithm.  Many configuration variables contribute in the accuracy of the genetic algorithms, e.g., probability of crossover and mutation operators. We had tuned such variables with trial and error, and we have utilized the tuned variables as described in Section \ref{sub:experimentdetails}. The population size and number of generations are also important variables in genetic algorithms. Figure \ref{fig:sensitivity} shows the sensitivity of the employed genetic algorithm to these two variables. The population size is set from 50 to 400 individuals, and the number of generations is varied from 50 to 200, with 50 intervals. The figure shows the average fitness of the proposed algorithm based on the NetDistance metric \cite{netdistance}. As the figure shows, the error of the proposed method is not meaningfully improved by increasing the number of generations. In other words, 50 generations is almost sufficient in our experiments for reaching the optimal results. Additionally, increasing the population size leads to better results but only until the population size reaches about 250 individuals. In summary, when the number of generations is more than 50 and the population size is more than 250, the evaluation results is relatively stable and less sensitive to these two configuration parameters.

\subsection{Discussion}
As illustrated in the evaluations, NetMix can adapt the topological features of the target network. In order to further investigate the capabilities of different methods in reproducing properties of the target networks, we can divide the considered network properties into two main categories: First, global (aggregate) properties and second, local properties and their distribution. The first category includes average clustering coefficient, transitivity, assortativity, and modularity, and the second category covers DDQC along with the distribution of degree, betweenness centrality, closeness centrality, eigenvector centrality, PageRank centrality, and local-clustering of the nodes. In the category of the global network properties, our proposed method (NetMix) outperforms all the baselines considerably. In the second category, NetMix and ABNG show similar accuracy for degree and closeness distribution, NetMix is better in DDQC, local-clustering, and eigenvector distribution. On the other hand, ABNG outperforms NetMix with respect to betweenness and PageRank distribution, mainly because in our experiments, no candidate process is defined for supporting these two properties. Fortunately, our proposed framework is extensible, and it is possible to add candidate processes for further supporting properties such as betweenness centrality, PageRank centrality, and other properties. 

When comparing NetMix with similar adaptive approaches, such as ABNG, we should also note the inherent advantages of NetMix. NetMix is naturally capable of synthesizing graphs with arbitrary size, perhaps different from the target graph. In this regard, NetMix follows a computationally inexpensive approach for generating large graphs, since it simply observes several features of the synthesized graphs in relatively small networks for training the process-probabilities and process-parameters during the proposed evolving genetic algorithm. On the other hand, size-dependent methods such as ABNG evolve a network with the same size as the target network. This approach is not practical for large target networks. Additionally, ABNG considers the whole distribution of several local properties, which results in time-consuming computations. Additionally, the distribution of the node properties is missing in some application domains, because we may access to only a limited properties of the target graph instead of the whole distribution of several properties. In other words, the needed information for applying ABNG is inaccessible or infeasible in many applications.   

\subsection{Details of the Experiments} \label{sub:experimentdetails}
It is worth noting some implementation details of the proposed method in our experiments, in order to make the reported results reproducible. We implemented the experiments using Python programming language and the NetworkX package. In configuring the GA, population size equals to 400 individuals, and a total of 200 generations are produced. The probability of applying mutation and crossover operators are set to 0.2 and 0.9 respectively. The ranges of chromosome features (described in Table \ref{tab:generativeparams}) are also preset as follows: $0 \leq P_{PA} + P_{TRA} + P_{MA} + P_{ADM} \leq 1$ , $ \frac{E}{N} - 2 \leq m \leq \frac{E}{N} + 2$, $ \frac{2 \times E}{N} - 2 \leq K \leq \frac{2 \times E}{N} + 2 $, $0 \leq P_{rewiring} \leq 1$, $0 \leq P_{copying} \leq 1$, $1 \leq N_{ADM} \leq \frac{N}{2}$, where E and N are number of desired edges and nodes respectively. The lower bound of $n$ is always set greater than or equal to the maximum possible value of $K$. The upper bound of $n$ is also set less than or equal to the number of desired nodes. It is worth noting that in our proposed method, the desired number of nodes in the synthesized network can be different from the number of nodes in the target network. Our proposed method generates the network in some consequent iterations. The number of performed iterations is equal to the desired number of nodes in the synthesized network, which can be different from the target network, divided by the $n$ value. The network formation starts from a $2 \times 2$ complete graph, and then adds $n$ nodes in each of the following iterations.

\section{Conclusion} \label{sec:conclusion}
In this paper, we proposed a novel method of network generation which is capable of adapting the target network. The proposed method, called NetMix, is actually a network model framework: NetMix automatically creates a network model based on the characteristics of the target network. In our proposed method, existing processes (such as transitive attachment or preferential attachment) are combined, which results in a mixed model for generating graphs that are topologically similar to the target network. We employed Genetic Algorithm in order to automatically find the best mixing configuration of the processes and the process parameters. The evaluations show the effectiveness of the proposed framework in reproducing topological features of both real and artificial networks. The experiments also verify that the proposed method outperforms baseline network models according to accuracy of different network properties in various network instances. Actually, the proposed method benefits from a simple extensible and architecture. NetMix is also a size-independent method, since it can generate networks similar to target graphs but with different number of nodes.

In summary, while a single existing network process (such as preferential attachment process of Barab\'{a}si-Albert model\cite{BAModel}) is unable to imitate topological features of all real-world networks, an appropriate adapted mixture of the existing network processes is capable of generating graphs similar to different kinds of real-world networks. As a future work, we will consider other network processes in our proposed framework to support broader range of network features. Additionally, we will investigate improving the GA fitness function. We will also apply the proposed method in different applications, particularly in biological network generation. 

\section*{Acknowledgement}
We are grateful to Viplove Arora for sharing the synthesized networks which were generated in the evaluation process of ABNG method \cite{arora2017action}.

\bibliography{NetCombin}

\end{document}